\documentclass[12pt]{iopart}

\usepackage{iopams}
\usepackage[dvips]{graphicx}

\begin{document}

\title{On the chemical composition of cosmic rays of highest energy}

\author{Grzegorz Wilk}
\address{The Andrzej So{\l}tan Institute for Nuclear Studies,
Ho\.{z}a 69, 00681, Warsaw, Poland} \ead{wilk@fuw.edu.pl}
\author{Zbigniew W\l odarczyk}
\address{Institute of Physics, Jan Kochanowski University,
\'Swi\c{e}tokrzyska 15, 25-406 Kielce, Poland}
\ead{zbigniew.wlodarczyk@ujk.kielce.pl}

\begin{abstract}
We present arguments aiming at reconciling apparently
contradictory results concerning the chemical composition of
cosmic rays of highest energy, coming recently from the Auger and
HiRes collaborations. In particular, we argue that the energy
dependence of the mean value and root mean square fluctuation of
shower maxima distributions observed by the Auger experiment are
not necessarily caused by the change of nuclear composition of
primary cosmic rays. They could also be caused by the change of
distribution of the first interaction point in the cascade. A new
observable, in which this influence is strongly suppressed, is
proposed and tested.
\end{abstract}


\maketitle

Recently two leading cosmic ray (CR) experiments, the Pierre Auger
Collaboration (Auger) \cite{Auger} and The High Resolution Fly's
Eye Collaboration (HiRes) \cite{HiRes} published their most recent
data on the depth of maxima of extensive air showers above
$10^{18}$ eV. Two apparently contradictory conclusions were
presented. Whereas the Auger collaboration cautiously concluded
that their data indicated a gradual increase of the average mass
of incoming CR with energy, HiRes stated that their data were
consistent with a predominantly protonic composition of cosmic
rays. These results started a vivid discussion \cite{PT}. In this
note, we propose a possible reconciliation of both results with an
indication that, perhaps, there is no need to introduce a heavy
(i.e., iron) component in the CR chemical composition. Namely, we
indicate that these results could also be due to the influence of
the distribution of the first interaction point in the cascade, at
least partially. We therefore propose and test a new observable,
cf. Eq. (\ref{eq:obs}), in which this influence is strongly
suppressed.

\begin{figure}[t]
\centerline{\includegraphics [width=12cm]{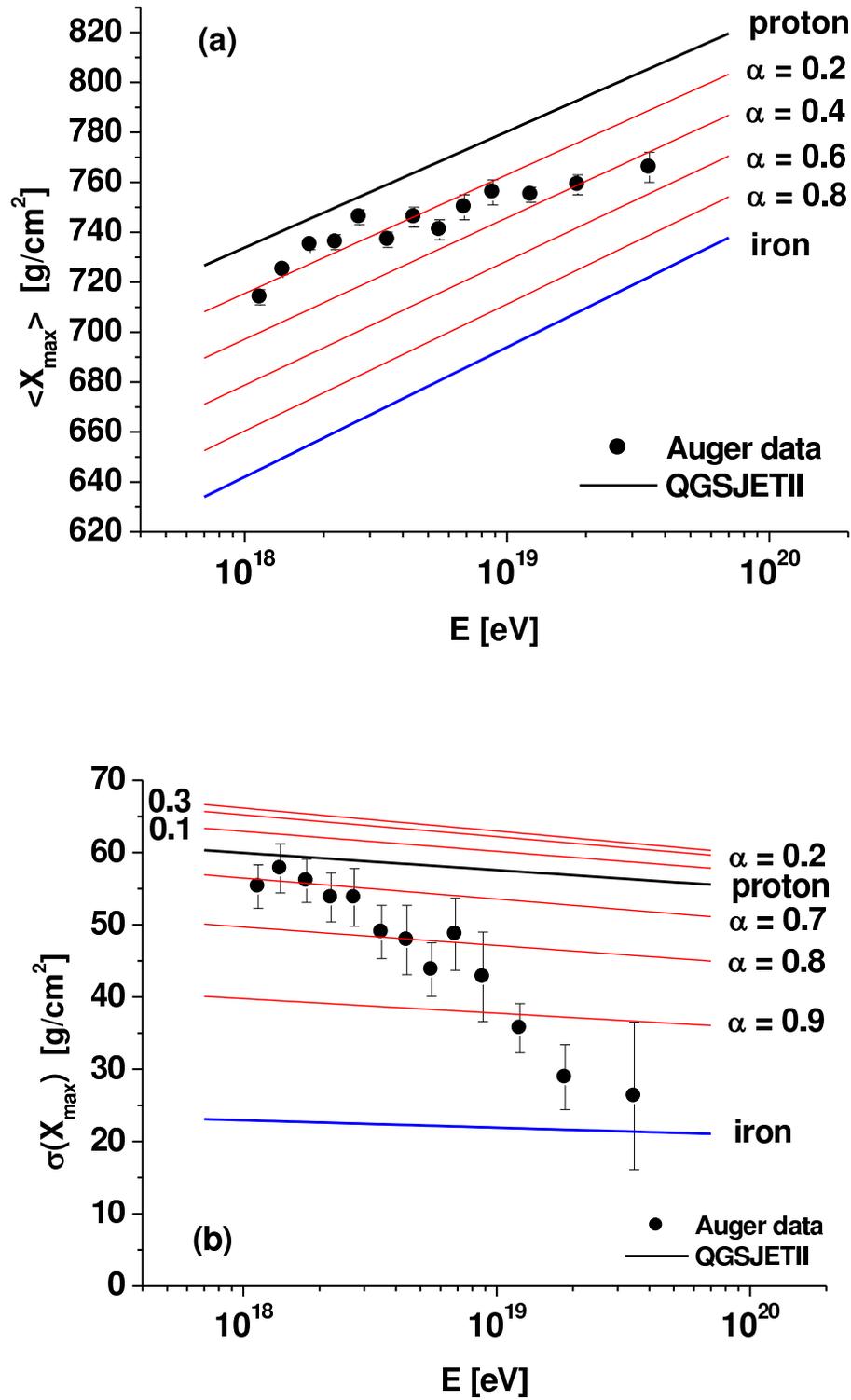}}
 \caption{(Color online) $\langle X_{max}\rangle$ and $\sigma \left(
           X_{max}\right)$ from \cite{Auger} compared with the QGSJETII model
           \cite{QGSJETII} using two components, protons and iron, cf. Eqs.
           (\ref{eq:Xmax}) and (\ref{eq:RMS}), respectively, with $\alpha$
           denoting the relative abundance of iron nuclei. }
 \label{fig1}
\end{figure}

\begin{figure}[h]
\centerline{\includegraphics [width=12cm]{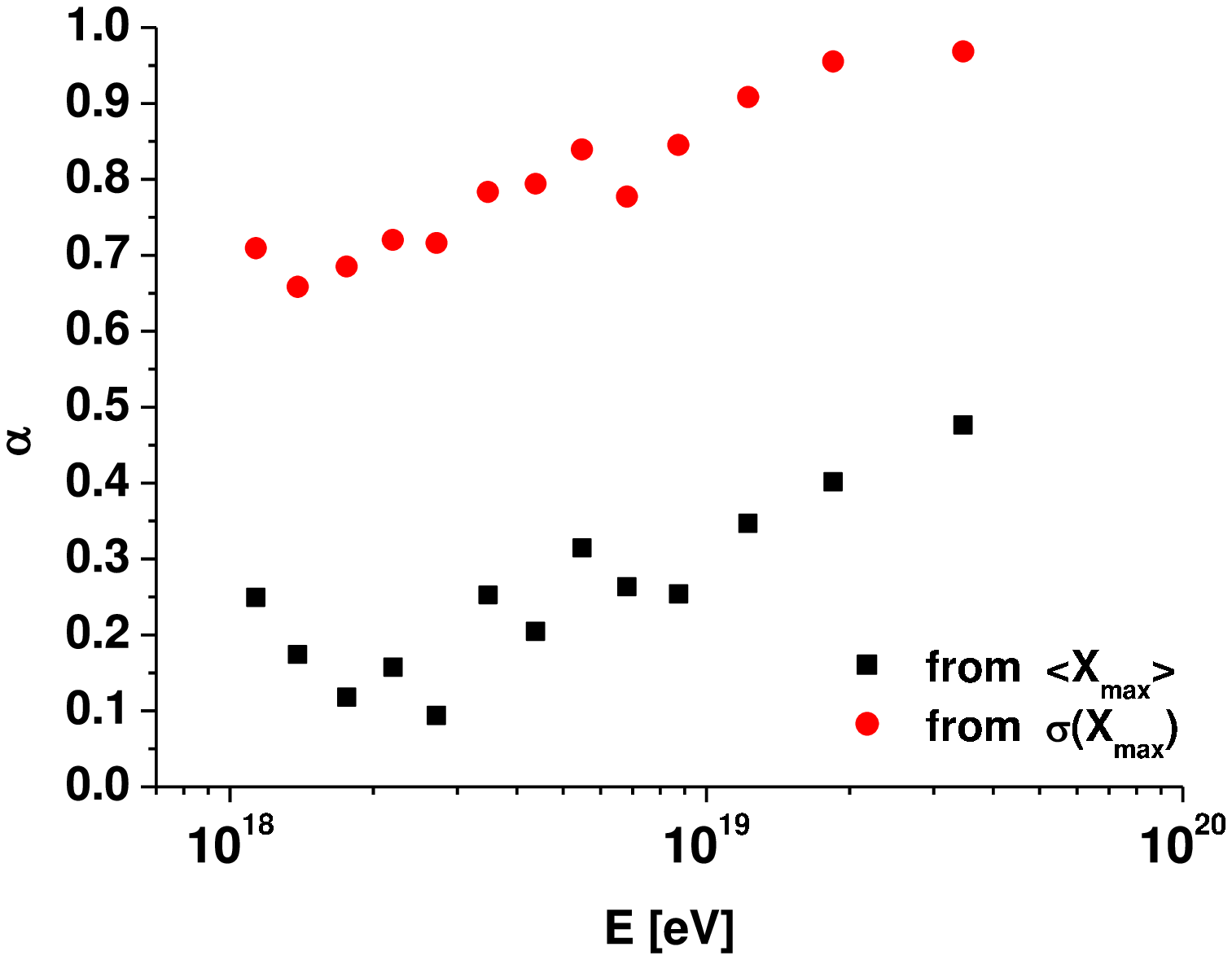}}
 \caption{(Color online) The energy dependence of relative abundance of iron in CR
           as extracted from $\langle X_{max}\rangle$ and $\sigma \left(
           X_{max}\right)$ shown in Fig. \ref{fig1} (in the frame of the QGSJETII
           model \cite{QGSJETII}).}
 \label{fig2}
\end{figure}

With increasing energy the Auger data \cite{Auger} show almost
monotonic changes in the chemical composition of CR changing from
proton to iron for two types of observables considered: the mean
depth of the maximum of the longitudinal development of air
showers, $\langle X_{max}\rangle$, and the shower-to-shower
fluctuations, the root mean square (rms) $\sigma \left(
X_{max}\right)$, see Fig. \ref{fig1} \footnote{In all figures
presented here we use both experimental data and model predictions
for pure proton and iron primaries following \cite{Auger,HiRes}.
Using information on their values of $\langle X\rangle_p$,
$\langle X\rangle_{Fe}$, $\sigma_p$ and $\sigma_{Fe}$, one can
deduce from Eqs.(\ref{eq:Xmax}) and (\ref{eq:RMS}), in an univocal
way, the corresponding values of observables of interest for a
given value of the parameter $\alpha$; in particular, the energy
dependencies of $\langle X_{max}\rangle$ and $\sigma\left(
X_{max}\right)$. The differences $\langle X_{max}\rangle -
\sigma\left( X_{max}\right)$ are evaluated directly from the
definition using the energy dependencies of $\langle
X_{max}\rangle$ and $\sigma \left( X_{max}\right)$ as given by the
models. Notice that Auger compares their data with pure
simulations, whereas HiRes quotes data including all detector
effects and compares them to the models {\it after} the detector
simulation. It means then that, unfortunately, both approaches
cannot be compared directly.}. For $\langle X_{max}\rangle$, a
such dependence can be interpreted by allowing for the presence of
two components in CR: iron, with relative abundance $\alpha$, and
protons, with relative abundance $1-\alpha$:
\begin{equation}
\langle X_{max}\rangle = (1 - \alpha)\langle X_{max}\rangle_p +
\alpha \langle X_{max}\rangle_{Fe}, \label{eq:Xmax}
\end{equation}
(where $\langle X_{max}\rangle_p$ and $\langle
X_{max}\rangle_{Fe}$ denote the mean depth of shower maxima for
the pure $p$ and $Fe$ CR's, respectively). However, the same
reasoning applied to $\sigma \left( \langle X_{max}\rangle
\right)$ lead to nonmonotonic dependence on $\alpha$ in this case
(seen as nonlinear spacing between lines corresponding to
different values of $\alpha$, cf. Fig. \ref{fig1} b),
\begin{equation}
\sigma^2 = (1 - \alpha)\sigma^2_p +\alpha \sigma^2_{Fe} + \alpha(1
- \alpha)\left( \langle X_{max}\rangle_p - \langle
X_{max}\rangle_{Fe} \right)^2. \label{eq:RMS}
\end{equation}
This has a maximum at
\begin{equation} \alpha = \frac{1}{2}\left[
1 - \frac{\sigma_p^2 - \sigma^2_{Fe}}{\left( \langle
X_{max}\rangle_p - \langle X_{max}\rangle_{Fe} \right)^2}\right].
\label{eq:maximum}
\end{equation}
This is seen in Fig. \ref{fig1}b, where one observes that adding
iron to protons results first (for small $\alpha$) in increased
fluctuations, and only for a quite large admixture of iron (large
$\alpha$) do they decrease towards the pure iron line. For this
reason, as seen in Fig. \ref{fig2}, for such a simple
parametrization, experimental data with similar energy behavior
lead to  quite different chemical compositions, ranging from
proton dominated for $\langle X_{max}\rangle$ to iron dominated
for $\sigma \left( X_{max}\right)$.

Among possible reasons of such a discrepancy, we shall concentrate
on problems connected with the development of the cascade, in
particular on the significance of the depth of the first
interaction. It is known (cf. \cite{A,R,U}) that most charged
particles in the shower are electrons and positrons with energies
near the critical energy $\varepsilon$ originated from the
electromagnetic subshowers initiated by photons from $\pi^0$
decay. The mean depth of the maximum of an electromagnetic shower
initiated by a photon with energy $E_{\gamma}$ is
\begin{equation}
\langle X^{em}_{max}\left( E_{\gamma} \right)\rangle = X_0 \ln
\left( \frac{E_{\gamma}}{\varepsilon} \right). \label{eq:Xem_max}
\end{equation}
Here $X_0 \approx 37$ g/cm$^2$ is the radiation length and
$\varepsilon = 81 $ MeV in the air. A nuclear-initiated shower
consists of a hadronic core which feeds the electromagnetic
component mainly through the production of $\pi^0$. Therefore, in
general, for an incident nucleus of mass $A$ (including protons
with $A=1$) and total energy $E$, the depth of the shower maximum
is given by
\begin{equation}
\langle X_{max}\rangle = \langle X^{em}_{max}\left[
\left(E/A\right)\left( K/\langle n\rangle\right)\right]\rangle +
\langle X_1\rangle, \label{eq:max}
\end{equation}
where $\langle X_1\rangle$ is the mean depth of the interaction
with maximal energy deposition into the shower (known also as the
depth of the first interaction), $K$ denotes inelasticity and
$\langle n\rangle$ is related to the multiplicity of secondaries
produced in the high-energy hadronic interactions in the cascade.
When the composition changes with energy, $\langle A\rangle$
depends on energy and $\langle X_{max}\rangle$ changes
accordingly. For primary nuclei with mass number $A$ and energy
$E$, the shower is, within a good approximation, simply equivalent
to a bundle of $A$ nucleons with energies $E/A$ each. In the case
of primary protons in the hadronic cascade, there is a hierarchy
of energies of secondary particles in each interaction and a
similar (approximately geometrical) hierarchy of interaction
energies in the cascade. In this case $\langle n\rangle$ has to be
understood as some kind of effective multiplicity without a
general straightforward definition. In addition, the inelasticity
$K$ can also change with energy \cite{KKK}.

Now, the probability of observing the first interaction in a
shower at a depth greater than $X$ is
\begin{equation}
P\left( X_1 > X\right) \propto \exp (-X/\lambda), \label{eq:L}
\end{equation}
where $\lambda$ denotes the interaction length (and is therefore
connected to the cross section, in our case $\lambda_{p-air} =
24160/\sigma_{p-air}$[g/cm$^2$] for cross-section given in [mb]).
It is tempting to use directly the exponential distributions of
showers with large $X_{max}$ to calculate $X_1$ (and the
proton-air cross section). However, this can be done only in the
case of a perfect correlation between $X_{max}$ and $X_1$. The
fluctuations existing in shower development modify such a
relation, leading to
\begin{equation}
P\left( X_{max} > X\right) \propto \exp( - X/\Lambda),
\label{eq:LL}
\end{equation}
where $\Lambda =k \lambda$, and $k$ accounts for the way the
energy dissipation takes place in the early stages of shower
evolution; it is particularly sensitive to the mean inelasticity
and its fluctuations. The factor $k$ depends mainly on the  way
the energy dissipation takes place in the early stages of shower
evolution and is particularly sensitive to the mean inelasticity
and its fluctuations. Small fluctuations in multiplicity and $K$
result in smaller $k$ \footnote{Assuming similar fluctuations in
multiplicity and inelasticity, a model predicting large $\langle
n\rangle$ of secondary particles leads to a smaller overall
fluctuations of the cumulative shower profile of secondaries,
i.e., to smaller factor $k$ \cite{A,U}.}.

In our case, in the absence of internal fluctuations, all showers
would develop between the first interaction point and the maximum
in the same amount of matter, $\Delta X=X_{max} - X_1$, showing
perfect correlation between the $X_{max}$ and $X_1$. This means
that their distributions should in this case have exactly the same
shape, but shifted by $\Delta X$, i.e., the slope of the $X_{max}$
distribution, $\Lambda$, should be equal to the mean interaction
length $\lambda$. This relation will, however, be affected by some
inevitable intrinsic fluctuations in shower development taking
place after the first interaction\footnote{A comprehensive studies
of the shower to shower fluctuations by means of Monte Carlo
simulations can be found in recent works \cite{MC_fluct,recent}.}.
In this case, because of fluctuations in $\Delta X$, the
correlation between $X_{max}$ and $X_1$ will diminish. Roughly,
one can write that
\begin{equation}
\sigma\left( X_{max}\right) \cong \sigma\left( X_1\right) + \xi[
\sigma( \Delta X )], \label{eq:sigma}
\end{equation}
where $\sigma \left( X_1\right) \propto \langle X_1\rangle$ and
the function $\xi (\sigma)$ describes the influence of shower
fluctuations after the first (main) interaction point. Notice that
for the probability distribution given by Eq.(\ref{eq:L}) the
fluctuation in $X_1$ is $\sigma\left( X_1\right) = \sqrt{Var\left(
X_1\right)} = \langle X_1\rangle = \lambda $. However, in the case
when $X_1$ is interpreted as the main interaction point (in which
the energy deposition to the shower is maximal) one would obtain a
gamma distribution (instead of Eq. (\ref{eq:L})) for which $\sigma
\left( X_1\right) = \sqrt{Var \left( X_1\right)} =\langle
X_1\rangle/\sqrt{\kappa}$, where $\kappa$ depends on the mean
inelasticity, $\langle K\rangle$, and determines in which of the
successive interactions of a projectile the energy deposition to
the shower is maximal\footnote{Numerically for $\langle K\rangle
\cong 0.7$ one has $\sigma \left( X_1\right) \cong 0.96 \langle
X_1\rangle$.}. To summarize, because of Eq. (\ref{eq:max}), where
$\langle X_{max}\rangle = \langle X^{em}_{max}\rangle + \langle
X_1\rangle $, we propose to use the following observable in which
the influence of fluctuations of the first interaction is strongly
suppressed,
\begin{eqnarray}
\langle X_{max}\rangle - \sigma\left( X_{max}\right) \cong \langle
X^{em}_{max}\left[ \left( E/A\right)( K/\langle n\rangle )
\right]\rangle - \xi[\sigma(\Delta X)] . \label{eq:obs}
\end{eqnarray}

\begin{figure}[h]
\centerline{\includegraphics [width=12cm]{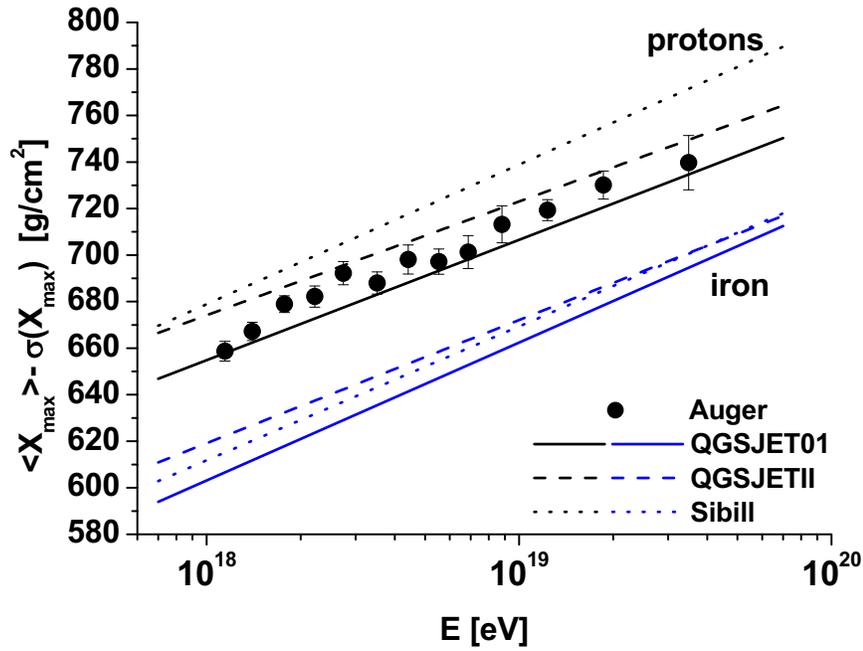}}
 \caption{(Color online) $\langle X_{max}\rangle
 - \sigma\left( X_{max}\right)$ as deduced from the Auger data
 and compared to different models \cite{Auger} for showers initiated
 by protons and iron. Note that experimental data prefer a proton composition.}
  \label{fig3}
\end{figure}

\begin{figure}[h]
\centerline{\includegraphics [width=12cm]{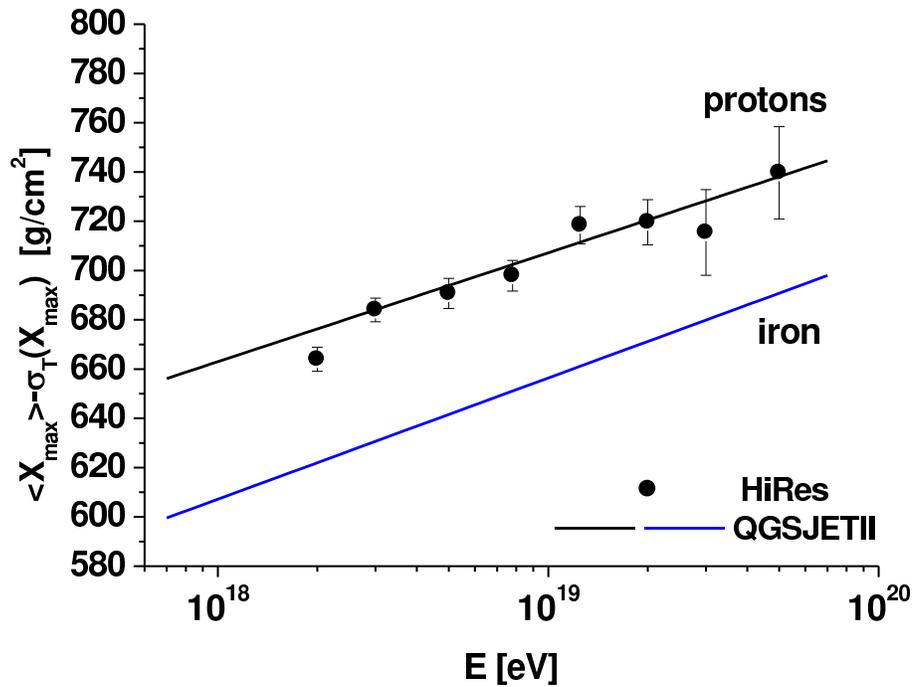}}
 \caption{(Color online) The same as in Fig. \ref{fig3}, but for the HiRes
 experimental data (in this case data are truncated at $2\sigma$) \cite{HiRes}. }
 \label{fig4}
\end{figure}

\begin{figure}[h]
\centerline{\includegraphics [width=12cm]{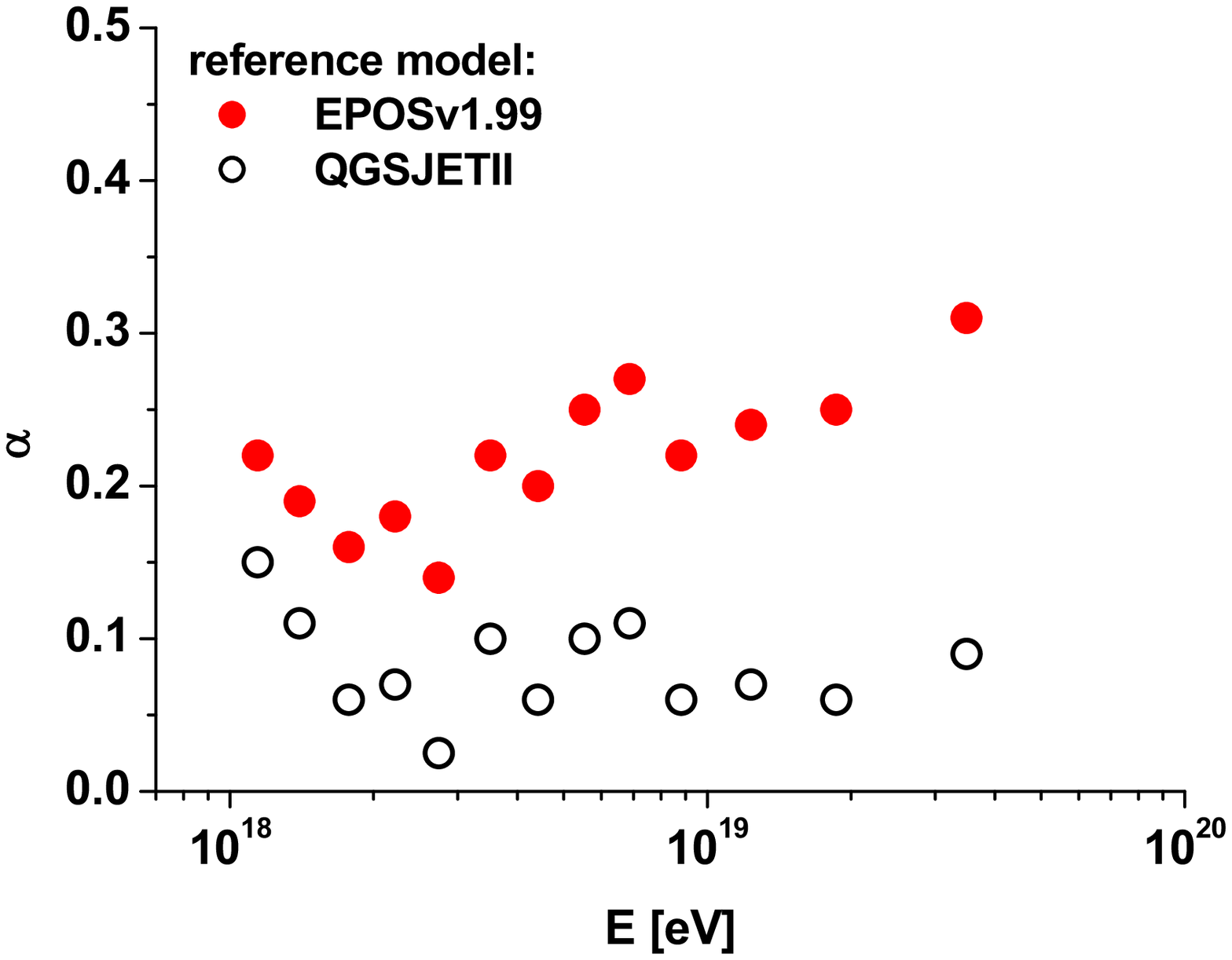}}
 \caption{(Color online) The energy dependence of the relative abundance
 of iron in CR as extracted from $\langle X_{max}\rangle -
\sigma\left( X_{max}\right)$ given by the Auger experiment and
shown in Fig. \ref{fig3}.
 }
 \label{fig5}
\end{figure}

To test this observable, we first plot its energy behavior in
Figs. \ref{fig3} and \ref{fig4} for the, respectively, Auger
\cite{Auger} and HiRes \cite{HiRes} data. Notice that now the
HiRes data, where the distribution of $X_{max}$ was truncated at
$2\sigma$ ($\sigma_T$ denotes truncated fluctuations), show
similar behavior as the Auger data. Notice also (cf. Fig.
\ref{fig3}) that $\langle X_{max}\rangle - \sigma\left(
X_{max}\right)$ given by Eq. (\ref{eq:obs}) still depends on
models of multiparticle production and is sensitive to the
chemical composition of CR ($p$ and $Fe$ initiated showers are
markedly different). Finally, Fig. 3 also tells us that the
chemical composition cannot be the origin of the observation by
Auger \cite{Auger} that $\langle X_{max}\rangle$ rises too slowly
with energy and approximates the expectation for primary $Fe$
nuclei. In fact, experimental data seem rather to be in fair
agreement with the hypothesis of a proton dominant composition of
the primary CR flux (assuming, of course, that the reference
models used are roughly correct). Within the toy model of primary
composition used before (with only two components: iron nuclei
with relative abundance $\alpha$ and protons with abundance $1 -
\alpha$, cf. Eqs. (\ref{eq:Xmax}) and (\ref{eq:RMS})) we can again
evaluate $\alpha$ as given by the Auger experiment but this time
from $\langle X_{max}\rangle - \sigma\left( X_{max}\right)$. The
result is shown in Fig. \ref{fig5}. For reference model QGSJETII,
the abundance of iron is roughly independent of energy ($\alpha
\approx 0.05 \div 0.1$ ) and even for the model EPOS v.1.99
\cite{NEW} leading to the maximal abundance of iron, $\alpha$
increases with energy rather slowly (remaining in the interval
$\alpha \approx 0.15 \div 0.3$ ).


To summarize this part, we learn from Fig. \ref{fig3} that the
main contribution to the energy dependence of $\langle
X_{max}\rangle$ and $\sigma\left( X_{max}\right)$ observed by
Auger \cite{Auger} comes from $\langle X_1\rangle$. This, however,
can be affected by two factors: the cross section $\sigma_{inel}$
and inelasticity $K$ (in fact, not only by its mean value $
\langle K\rangle$, but also by its distribution). Roughly,
$\langle X_1\rangle = \lambda \cdot \kappa$, where $\kappa$
determines in which of the successive interactions of a projectile
the energy deposition to the shower is maximal. For a uniform
inelasticity distribution (in the maximal possible interval for a
given $\langle K\rangle$), one has $\kappa \simeq 1 + 1.85(0.75 -
\langle K\rangle)$.  As shown recently in \cite{Kodama}, if gluon
saturation occurs in the nuclear surface region then
$\sigma_{p-air}$ at $E > 10^{18}$ eV increases more rapidly with
incident energy than is usually estimated. Although in
\cite{KKK,OurK} we have argued for an overall decrease with energy
of the inelasticity $K$, its increase at energies $E \approx
10^{18}$ eV is by no means excluded \footnote{Recently the role of
inelasticity in high energy CR was discussed in \cite{DdD} using
the percolation theory approach.}. Both possibilities require an
abrupt onset of some "new physics" beyond the standard model,
which would be difficult to accept. It is worth mentioning as an
example the elongation rate, which in the case of Eq.
(\ref{eq:max}) is given by \cite{A,el}
\begin{equation}
D_{10} = \frac{d\langle X_{max}\rangle}{d \log E} =
\frac{X_0}{\log e}\left[ 1 - \frac{d \log \left( A\langle
n\rangle/K\right)}{d\log E} \right] + \frac{d \langle
X_1\rangle}{d \log E}. \label{eq:D}
\end{equation}
As reported in \cite{Auger_el}, Auger observes the apparently
abrupt change in $D_{10}$ at energy $\approx 2\cdot 10^{18}$ eV,
which could signal some new physics. If we denote $D^{\star}_{10}
= d\left[ \langle X_{max}\rangle - \sigma \left( X_{max}
\right)\right]/d \log E$, then one expects here something of the
order of $D^{\star}_{10} - D_{10} = - d\sigma \left(
X_{max}\right)/d \log E \cong  3.5$ g$\cdot$cm$^{-2}$, depending
on the increase in cross section adopted (chosen from existing
models and predictions). However, the Auger data above $10^{18}$
provide the value $22$ g$\cdot$cm$^{-2}$. It can be shown that
such a large value leads to a very strong energy dependence of the
cross section, $d \sigma_{p-air}/d \log E \cong 0.48
\sigma_{p-air}$, which seems at the moment to be very unrealistic
and contradicts even the scenario of gluon saturation on the
nuclear surface recently proposed in \cite{Kodama}. A more
detailed discussion of this problem is outside the scope of this
note. On the other hand, the center of mass collision energies of
the order of few hundred TeV observed here are well beyond those
to be studied in the foreseeable future at LHC. This means that CR
are, most probably, the only future source of information on the
properties of interactions at these energies and surprises should
not be ruled out {\it a priori}.

\begin{figure}[h]
\vspace{5mm} \centerline{\includegraphics [width=12cm]{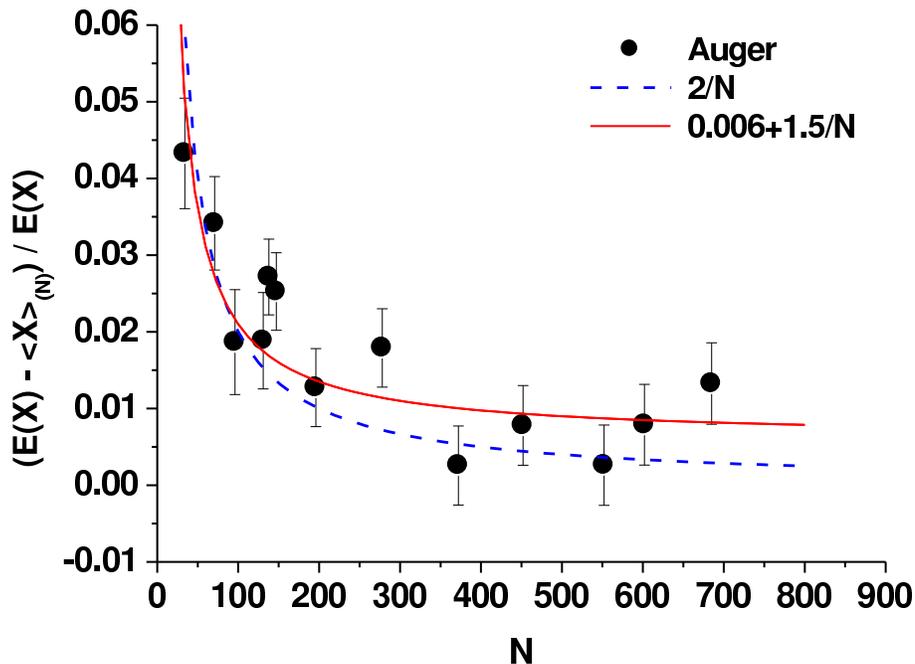}}
 \caption{(Color online) Illustration of the effect of statistics
 as given by Eq. (\ref{eq:comp}) with
 $\langle X\rangle_{(\infty)} = E(X) $ assumed to be the same as the mean
 value given by the QGSJET01 model \cite{QGSJET01}, see text for details.}
 \label{fig6}
\end{figure}

\begin{figure}[h]
\vspace{5mm} \centerline{\includegraphics [width=12cm]{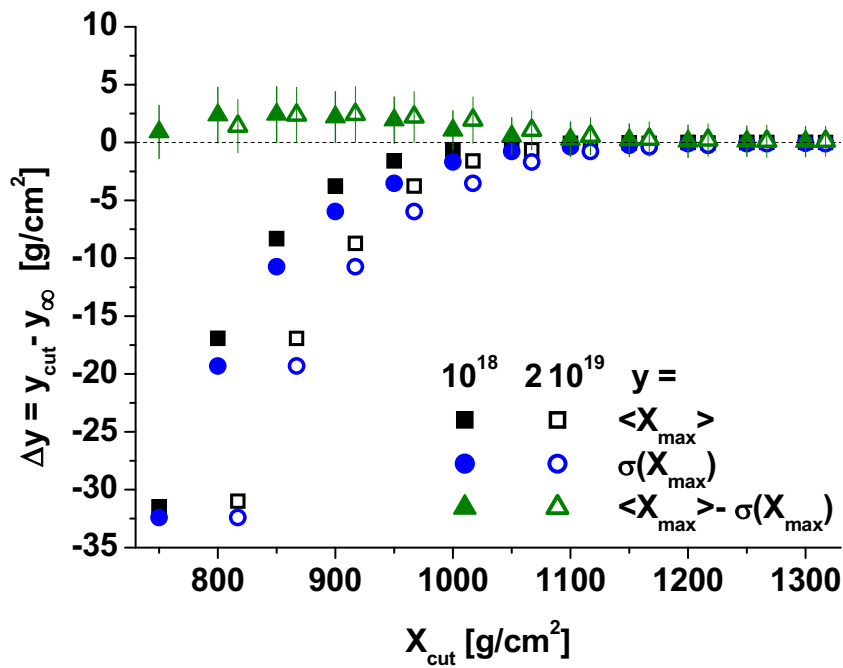}}
 \caption{(Color online) Difference of the observable
           $y_{cut}$, evaluated from the $X_{max}$ distribution truncated at
           $X_{cut}$, and $y_{\infty}$, evaluated from the unbiased
           distribution (given by QGSJET01 model for primary protons
           with energy $10^{18}$ eV and $2\cdot 10^{19}$ eV).
           When energy increases, the observed dependence shifts
           towards the higher $X_{cut}$, proportionally to the increase of
           $\langle X_{max}\rangle$ with energy, i.e., the dependence
           of $\Delta y$ on $X_{cut} - \langle X_{max}\rangle_{\infty}$
           remains roughly the same for all primary energies.}
 \label{fig7}
\end{figure}

As mentioned above the Hires data (which are truncated at
$2\sigma$) \cite{HiRes} show similar energy dependence for
$\langle X_{max}\rangle - \sigma \left( X_{max}\right)$ as the
Auger data, c.f. Fig. \ref{fig4}. This indicates that, in both
cases, the crucial factors are the tails of the $X_{max}$
distributions. For a small sample, the values of $X_{max}$ near
the maximum of the distribution are preferred, and the estimated
mean value $\langle X_{max}\rangle$ differs from the expected
value, $E\left( X_{max}\right)$. To investigate whether the effect
observed by Auger could be connected to small statistics, notice
that, because the distribution of distances of the first
interaction is $\exp(-X/\lambda)$, when calculating $\langle
X\rangle = \frac{1}{N}\sum^N_i X_i$ one encounters $S =
\sum^N_iX_i$ which has a gamma distribution,
\begin{equation}
p(S) = \frac{S^{N-1}}{\lambda^N\Gamma(N)}\exp\left(
-\frac{S}{\lambda}\right). \label{eq:P(S)}
\end{equation}
For small samples one in fact observes the most probable values,
which for a gamma distribution is equal to $S^{(max)} = \lambda
(N-1)$. This means that we can expect that, for a sample
consisting with $N$ elements one has
\begin{equation}
\langle X\rangle_{(N)} = \frac{1}{N} \sum^N_i X_i \approx
\frac{1}{N} S^{(max)} = \lambda \left( 1 - \frac{1}{N} \right),
\label{eq:samplemean}
\end{equation}
whereas the true expectation values for the exponential
distribution (obtained for $N \rightarrow \infty$) is $E(X) =
\lambda$. Therefore, for a sample of $N$ elements the estimator
$\langle X\rangle_N$ is biased by a value of the order of
\begin{equation}
E(X) - \langle X\rangle_{(N)} \approx \frac{\lambda}{N}.
\label{eq:bies}
\end{equation}
In Fig. \ref{fig6} we compare
\begin{equation}
\frac{ \langle X\rangle_{(\infty)} - \langle X\rangle_{(N)}}{
\langle X\rangle_{(\infty)}}  = 1 - \frac{ \langle X\rangle_{(N)}
}{ \langle X\rangle_{(\infty)}} \propto \frac{1}{N}
\label{eq:comp}
\end{equation}
with data of Auger, here $\langle X\rangle_{(\infty)}= E(X)$
denotes the value given by the model. Because the reference model
(here QGSJET01) does not exactly describe the experimental data
(even in the lower energy region, i.e. for large values of $N$) we
use here the simple formula $a+b/N$ to describe the dependence on
$N$. To further illustrate the significance of tails of $X_{max}$
distributions, we examine the truncated $X_{max}$ distribution in
the interval $(0, X_{cut})$, cf., Fig. \ref{fig7} where $\langle
X_{max}\rangle$, $\sigma\left( X_{max}\right)$ and $\langle
X_{max}\rangle - \sigma\left( X_{max}\right)$, evaluated for
different $X_{cut}$, are presented. Notice that $\langle
X_{max}\rangle$ and $\sigma \left( X_{max}\right)$ are strongly
dependent on the value of $X_{cut}$. On the other hand, their
difference introduced in Eq. (\ref{eq:obs}) above, $\langle
X_{max}\rangle - \sigma \left(X_{max}\right)$, is rather
insensitive to the cutting procedure used. Therefore, the possible
biasses of the tail of the $X_{max}$ distribution do not influence
this observable.

In conclusion, we argue that the spectacular energy dependence of
the shower maxima distribution reported by the Auger collaboration
\cite{Auger} is not necessarily (or not only) due to the changes
of chemical composition of primary cosmic rays. The observed
effect (or, at least, a substantial part of it) seems rather to be
caused by the unexpected changes of the depth of first
interaction, $X_1$. However, the energy dependence of $\langle
X_1\rangle$ can be affected by a rapid increase of cross section
and/or increase of inelasticity in energies above $2\cdot 10^{18}$
eV. Both possibilities require an abrupt onset of some "new
physics" in this energy region and are therefore questionable. The
HiRes data \cite{HiRes}, where the $X_{max}$ distribution was
truncated and, after that operation, is consistent with the proton
spectrum, brings in the possible role of biases of the $X_{max}$
distribution indicating that the ways of analyzing CR data of
highest energy still remains an open problem. We argue that it
would be highly desirable to analyze the observable $\langle
X_{max}\rangle - \sigma \left( X_{max}\right)$ (cf. Eq.
(\ref{eq:obs})) in which fluctuations of the depth of the first
interaction, as well as the possible biases of the tail of
$X_{max}$ distribution, are strongly suppressed. This observable
still depends on the model of multiparticle production and is
sensitive to the chemical composition of the primary CR.

Summarizing, though the problem of the chemical composition of CR
seem still unresolved, we expect that the large spread observed in
the results can find an explanation by comparing the different
sensitivities to the composition of CR in various observables. A
deeper understanding, both of the hadronic interaction models and
of the systematics of data analysis, which could bias the results,
is also needed. Such a detailed analysis, with detailed simulation
studies, is, however, outside the scope of this paper.

\section*{Acknowledgment}

Partial support (GW) of the Ministry of Science and Higher
Education under contract DPN/N97/CERN/2009 is gratefully
acknowledged. We would like to thank warmly dr Eryk Infeld for
reading this manuscript.

\end{document}